\documentstyle[12pt,fleqn,cite]{article}

\def\be{\begin{equation}}
\def\ee{\end{equation}}
\def\bea{\begin{eqnarray}}
\def\eea{\end{eqnarray}}

\begin{document}
\begin{titlepage}
\begin{center}
\hfill hep-th/9812081\\
\hfill CPHT/694.1298\\
\hfill IP/BBSR/98-38\\
\hfill MRI-PHY/P981272\\

\vskip .2in

{\Large \bf String Multiplets from an Invariant 2-Brane}
\vskip .5in

{\bf Sandip Bhattacharyya$^*$, Alok Kumar$^{*, \dagger}$ and 
Subir Mukhopadhyay$^{\dagger\dagger}$\\
\vskip .1in
{\em  * Institute of Physics
{\footnote{email:sandip,kumar@iopb.res.in,subirm@mri.ernet.in}},\\
Bhubaneswar 751 005, INDIA}}
\vskip .7cm
{\em $\dagger$ Centre de Physique Theorique,\\
Ecole Polytechnique,\\
91128 Palaiseau, France}
\vskip .7cm
{\em $\dagger\dagger$ Mehta Research Institute,\\
Allahabad 221 506, INDIA}
\end{center}

\begin{center} {\bf ABSTRACT}
\end{center}
\begin{quotation}\noindent
\baselineskip 10pt
We study a three-dimensional gauge theory obtained from the 
dimensional reduction of a D4-brane worldvolume theory
in the background of space-time moduli. 
An $SL(3)$ symmetry  in this theory, 
which acts on fields as well as coupling constants, is 
identified. By comparing the energies with the string tensions, 
we show that certain 1/2 supersymmetric classical solutions of this theory
can be identified as $SL(3,Z)$ multiplets  of type II strings in 
eight dimensions. Results are then generalized to the non-linear 
Born-Infeld action. We also discuss the possibility of 
1/8 BPS states in this theory and their representations
in terms of string networks.

\end{quotation}
\vskip .2in
CPHT/694.1298\\
IP/BBSR/98-38\\
MRI-PHY/981272\\
December 1998\\
\end{titlepage}
\vfill
\eject



BPS soliton solutions of the worldvolume gauge 
theories\cite{calmal,hashi1,gibbons,gauntlett,prasanta} have recently
been a subject of intensive discussions. 
These solutions describe the brane configurations, whose 
existence on the basis of supersymmetry and charge conservation was 
argued earlier. 
The worldvolume description is also suitable
for discussing their dynamics\cite{bachas}. 
The brane dynamics and their correspondence to higher 
dimensional conformal theories have also been pointed out
\cite{cal2}. In this paper we discuss similar solutions of certain
three-dimensional Born-Infeld 
and Yang-Mills theories. Our work 
has been motivated by the recent progress in understanding the
$1/4$ BPS states of N=4 supersymmetric gauge theory in 
four dimensions\cite{bergmann} 
through string networks\cite{net,net2} involving 
$(p, q)$ strings\cite{sl2} of ten-dimensional type IIB string theory. 

It has been shown that
the string junctions and networks of ten-dimensional 
type IIB theories can also be 
identified as solutions of N=4 gauge theories in four 
dimensions\cite{hashi2,oku,lee}. 
The starting point of such studies is the D3-brane picture of 
$N=4$ gauge theories. More precisely, certain
``spike-solutions'' of 3-brane gauge and Born-Infeld theoies can 
be interpreted as fundamental and D-string solutions. In particular,
a fundamental string corresponds to a point electric-charge
and a D-string to a point magnetic charge 
solution\cite{calmal}. However to 
maintain 1/2 supersymmetry, one simultaneously has to turn on 
Higgs fields as well. The Higgs structure gives 
rise to the string interpretation, as the energy density of the
solution  is proportional to the value of the Higgs, 
interpreted as the string coordinate and the  
proportionality constant defines the string tension. 
The full nonlinear Born-Infeld (B-I) theory also possess 
these solutions and their energy remains
unchanged in the full theory, confirming their BPS nature. 

To generalize this idea to obtain the 1/4 supersymmetric BPS 
solutions corresponding to the 3-string junction of the 
type IIB theory\cite{bergmann},
one studies the situation when several 3-branes are present. 
The fact that the mass of the 3-string junction coincides
with that of the 1/4 supersymmetric dyon solutions can be checked 
directly from the tensions of the respective strings which 
form the junction\cite{bergmann,net}. 
However to directly obtain these states as
classical solutions in a gauge theory, one now studies the 
$SU(3)$ or higher $SU(N)$ non-abelian generalizations.
The Higgs structure of the solution now becomes more complicated.
It however admits the solutions which asymptotically has the
structure of a 3-prong string or even more general networks.
 
A central role in the above 
identification is played by the $SL(2, Z)$ S-duality symmetry 
of the ten dimensional type IIB string theory whose action on the
invariant 3-brane gives rise to the duality symmetry of the 
four dimensional Yang-Mills.  In this paper we present a similar
role for an $SL(3)$ duality symmetry of the type II string theory in 
eight dimensions.

It was shown by the present authors\cite{bhatt} that
string networks can be generalized to the strings carrying 
general U-duality charges. In particular for the $SL(3, Z)$ 
\cite{leung,bhatt,luroy} symmetry in eight dimensions, 
the networks preserve 1/8
supersymemtry. It is then of interest to examine how such 
states can arise in gauge theories. It is clear from 
the nature of the supersymmetry
of these states, that they can be of relevance only in three or lower
dimensions. In this context, as a natural generalization of the
3-brane gauge theories, we now study the
worldvolume theory of certain 
2-branes having an $SL(3)$ invariance. Such 2-branes,
also referred as U2-branes,
exist in the eight dimensional type II string theories and  
transform as a $({\bf 1, 2})$ representation of the $SL(3)\times SL(2)$
duality group. However, in this paper we will concentrate on the
$SL(3)$ part  only,
which is an invariance of such a 2-brane theory.
Although following related developments, it should be
possible to write down a membrane 
action taking into account the $SL(2)$ symmetry as 
well\cite{christof}. 
 
Extended duality symmetries of gauge 
theories\cite{taylor,verl2} have also 
been discussed in the context of Matrix theories for their
identifications with string theories in various dimensions. Unlike 
these cases, however, in the examples presented here,
only some of the space dimensions in a gauge theory action 
are compactified. As a result there are 
possibilites of static space-dependent solutions. In particular,
the string solutions we will discuss correspond to spherically symmetric 
solution in three dimensions.

The 1/2 supersymemtric solutions of 3-dimensional
theories, including the ones with a structure of a periodic array of
BPS monopoles in four dimensions,
have been discussed earlier in the context of gauge and
heterotic string theories\cite{sen3d}. 
It has been observed that unlike the 
higher dimensional solutions, the 3-dimensional ones have a singularity not
only in the ultraviolet, but in the infrared regime as well\cite{sen3d}. 
{ This has been related to the physical effect of 
a logarithmic confinement in 
the 3D theory\cite{seiberg}}. However in a classical theory, 
the possiblity of removing this 
singularity by a modification of the solution, similar to the one for
a ten-dimensional 7-brane solution as proposed by F-theory\cite{vafa},
in the infrared or long-distance regime
has also been suggested\cite{sen3d}, 
although no concrete model has been proposed where such singularities
are absent. As a result, the BPS solutions are considered with an 
infrared cut-off. We will make similar assumptions for discussing the
3-D solutions in this paper as well. 
In the context of the world volume gauge
theories, this kind of IR divergences
are present in the M theory lifts. However,
in those cases one can define some 
regularization using infra-red cut-off.

In this paper we first write down a
(linearized) action describing 
certain  U2-branes (membrane), namely the 2-branes that are  
representations of the $U$-duality group. This action arises 
from a D4-brane action by a toroidal compactification.
The $SL(3)$ duality symmetry of this theory can be described by 
its action on three scalar fields.
Two of these arise from the dimensional reduction of 
a five-dimensional gauge field and third scalar field is dual to 
a $U(1)$ gauge field in three dimensions. These scalar 
fields, or the 
corresponding dual vectors, define the charges in three dimensions
which transform as a triplet under $SL(3)$. One of these has 
a direct interpretation as the electric charge in three dimensions. 
Two others arise due to the compactification of the 5-dimensional vector 
fields. The $SL(3)$ invariance of the action also has a natural interpretation 
from the point of view of an M5-brane action which in turn follows from 
the correspondence between the D4 and M5-branes. It should also be 
mentioned that, in addition, appropriate reduction of a 
D2 or M2-brane actions can also 
give an $SL(3)$ invariant action. We however concentrate on the D4
case here.

To study the classical solutions and their energies,
we also write down the corresponding Hamiltonian. 
The $SL(3)$ invariance of the worlvolume theory is 
manifest in the form of the Hamiltonian as well. 
The Gauss-law constraint and other equations of motion allow
simple solutions in terms of two dimensional Green functions
with $1/2$ supersymmetry and non-zero $SL(3)$ charges.  
We then observe that the energy density of the solutions with 
non-zero electric charge can be identified  with the string tension of 
a $(1, 0, 0)$-string of the eight dimensional type II theory. 
Similarly the solutions with other two charges 
correspond to the $(0, 1, 0)$ and $(0, 0, 1)$ strings of the
eight dimensional type II theory. 
More general solutions with $(p, q, r)$ charges can be 
obtained as a rotation among these ones. 
We then consider the Born-Infeld action and once again obtain 
similar solutions. The Gauss-law constraint once again gives the 
Green-function equation through which explicit solutions are
obtained. As expected the energies of the solutions are identical 
in the linear as well as the non-linear Born-Infeld theories. 
Finally, we discuss the status of states with $1/8$ supersymmetry
in 3-d gauge theories through a non-abelian generalization of our
action. However their full solution is left as a future exercise.

We now start by writing down the classical action for the
U2-brane in eight dimensions. This can be obtained by 
compactifying the D4-brane action of the
ten dimensional IIA theory on a $T^2$.
In the linear approximation, 
the gauge and scalar kinetic terms  can be written
as the reduction of an N=1, D=10 super Yang-Mills
theory to $(2+1)$ dimension. One however has to take into
account the Wess-Zumino terms of the D4-brane theory as well. The
bosonic degrees of freedom in the three dimensional case 
contains a vector field $A_\mu$ and 7 scalars. Due to the 
compactification on a $T^2$, two of the scalar fields, $(\phi_m,
m=1,2)$ coming as internal components of a D4-brane gauge field
$A_{\bar{\mu}}$ in five dimensions,
take values on a compact torus while
the rest five, $X^I (I=1,..,5)$, remain non-compact. The fact that in
three dimensions the gauge field is equivalent to
a compact scalar field implies the theory
has an $SL(3)$ invariance under which all
three compact scalar fields transform into 
each other.

Now, the linearized form of the 
bosonic part of the D4-brane action 
in Eulidean space\cite{christof} has a form
$S\equiv I + I_S$, where:
\begin{equation}
I = -\int d^5 x \left( {1\over 4 g} {\cal F}_{\bar{\mu}\bar{\nu}}
        {\cal F}^{\bar{\mu}\bar{\nu}} 
+ {i \over 4}\epsilon^{\bar{\mu}\bar{\nu}\bar{\rho}
\bar{\alpha}\bar{\beta}}C_{\bar{\mu}}{\cal F}_{\bar{\nu} \bar{\rho}}
{\cal F}_{\bar{\alpha} \bar{\beta}} 
+ i \epsilon^{\bar{\mu}\bar{\nu}\bar{\rho}
\bar{\alpha}\bar{\beta}}C_{\bar{\mu}\bar{\nu} \bar{\rho}}
{\cal F}_{\bar{\alpha} \bar{\beta}} \right), 
\end{equation}
and $(\bar{\mu}, \bar{\nu} = 0,..,4)$ are the 5-dimensional space-time
indices and ${\cal F}_{\bar{\mu} \bar{\nu}} = 
F_{\bar{\mu} \bar{\nu}} - B_{\bar{\mu} \bar{\nu}}$ is an appropriate 
field strength\cite{bachas}. 
$I_S$ is the kintetic energy term for the five scalars
(by setting the background gauge fields from the NS-NS 2nd rank 
antisymmetric tensor to zero):
\begin{equation}
I_S = -{1\over 2 g}
\int d^5x \partial_{\bar{\mu}} X^I \partial^{\bar{\mu}}X^I
\end{equation}
and non-compact scalars
$X^I$ are the singlets of the $SL(3)$ group that we consider below. 
This fact is also consistent with the M5-brane point of view where
non-compact coordinates represent the transverse directions to the
brane and are therefore orthogonal to the three wrapped directions.

While reducing the action to three dimensions we make several 
assumptions about the space-time backgrounds. First, since our
interest is mainly in the $SL(3)$ part of the eight dimensional 
$U$-duality group, we set the $SL(2)/SO(2)$ moduli\cite{luroy}
to zero. This in particular implies that components 
$B_{m n} = 0$ for the antisymmetric tensor fields.
In addition we assume that there are no
vector field backgrounds coming from the compactification, 
namely $g_{\mu m}$, $C_{\mu m n} = 0$ and the 2nd-rank 
antisymmetric tensors coming from the compactification of the 
NS-NS as well as RR-forms are absent, namely 
$B_{\mu \nu}$ = $C_{\mu \nu m} = 0$. The restricted 
space-time backgrounds therefore 
correspond to the 2-branes being 
studied  in the flat eight-dimensional space. 
As in the case of D3-branes, such
backgrounds correspond to standard gauge theories
in three dimensions. However it is certainly of interest to study 
the gauge theory corresponding to the branes in non-trivial
space-time backgrounds as well\cite{cal2}. 
For example, this may be of intetest for 
$N=2$ gauge theory in four dimensions, where D3-branes are 
studied  in 7-brane backgrounds. In our case, 
only nontrivial space-time 
backgrounds therefore are the $SL(3)/SO(3)$ 
moduli, giving rise to certain 
coupling constants in the gauge theory.  With these simplifications,
the full lower dimensional action has a form:
\begin{equation}
I = - A. \int d^3 x \left[ {1\over 4 g} F_{\mu \nu}^2 + 
        {1\over 2 g} g^{m n}\partial_{\mu}
        \phi_m \partial^{\mu}\phi_n +
        i  \epsilon^{m n}\epsilon^{\mu \nu \rho}
        C_m \partial_{\mu} \phi_n F_{\nu \rho} \right]
                        \label{classical-action}
\end{equation}
where $A$ is the area of the compactified 2-dimensional space.
The manifest $SL(3)$ invariant form of this action is obtained 
through dualization of the vector field into a compact scalar
by adding the auxiliary antisymmetric tensor $b_{\mu \nu}$
to $F_{\mu \nu}$. The action then has an extra gauge invariance which 
can be used for eliminating the original gauge field 
$A_{\mu}$. This action has a form: 
\begin{eqnarray}
\tilde{I} & = & - A \int d^3x \left[ {1\over 4 g} 
        (F_{\mu \nu} - b_{\mu \nu})^2
        + {1\over 2 g} g^{m n}\partial_{\mu}\phi_m
        \partial^{\mu}\phi_n            \right. \cr 
        & & + i \epsilon^{m n}\epsilon^{\mu \nu \rho}
        C_m \partial_{\rho}\phi_n (F_{\mu \nu} - b_{\mu \nu})   
        + i \alpha \epsilon^{\mu \nu \rho} 
        b_{\mu \nu} \partial_{\rho}\sigma               
\left.                                  \right]
\end{eqnarray}
where the constant coefficient of $b_{\mu \nu}$ in the last term
is fixed by requiring 
a proper $SL(3)$ matrix after dualization. 
The dual action is now obtained by setting $F_{\mu \nu} = 0$ and 
integrating out $b_{\mu \nu}$:
\begin{equation}
\tilde{I} = - A \int d^3x \left[  {g} 
        ( \alpha \partial_{\mu}\sigma -  \epsilon^{m n}
        C_m \partial_{\mu} \phi_n)^2 
        + {1\over 2 g}g^{m n}\partial_{\mu}\phi_m 
        \partial^{\mu}\phi_n
        \right].                \label{dualaction}
\end{equation}

A similar reduction of a four dimensional gauge  
action leads to a 3D gauge theory\cite{seiberg} which has been used for  
interpolating between the four-dimensional 
$N=2$ and three-dimensional $N=4$ gauge theory. 
In the present case also, 
the three-dimensional theory has certain duality symmetries
which originate from the compactification and 
act on the coupling constants as well. As a result they are not the 
symmetries of the action, for fixed couplings, 
and instead relate the particle spectrum at
different couplings. 
To find the action of this $SL(3)$ symmetry,  
we write $\tilde{I}$ in eqn.(\ref{dualaction}) in a matrix form:
\begin{equation}
        \tilde{I} = - A\int d^3 x \partial_{\mu} \omega^T {\cal M}
        \partial^{\mu}\omega,
\end{equation}
where 
\begin{eqnarray}
\omega \equiv \pmatrix{\sigma \cr
        \phi_1 \cr \phi_2},
\end{eqnarray}
is a $3\times 1$ column vector and 
${\cal M}$ is a matrix written as:
\begin{eqnarray}
{\cal M} = \pmatrix{  {g\alpha^2} & - g \alpha  
                \epsilon^{p n}C_p \cr
        - g \alpha \epsilon^{p m}C_p & 
        {1\over 2 g}g^{m n} +  g \epsilon^{p m}
        \epsilon^{q n} C_p C_q}.
                \label{sl3-matrix}
\end{eqnarray}

The ${\cal M}$ in eqn. (\ref{sl3-matrix}) is a real symmetric matrix of 
unit determinant and  defines a coset structure 
$SL(3)/SO(3)$. The $SL(3)$ action on the fields and couplings 
are then given as:
\begin{equation}
\omega \rightarrow \Lambda \omega,\>\>\>
{\cal M} \rightarrow \Lambda^{-1 T} {\cal M} \Lambda^{-1}.
\end{equation}

We have therefore written down the action of a non-compact
$SL(3)$ symmetry in a compactified D4-brane worldvolume theory.
As mentioned earlier, the presence of such duality symmetries in 
compactified gauge theories are expected from other considerations
like Matrix theory. Extra quantum 
numbers giving rise to a representation of a larger duality symmetry
arise from the periodicity properties of the D4-brane gauge 
fields along compact directions.
However, as emphasized earlier, our aim in the
paper is to obtain explicit classical solutions of the compactified
theory and to identify them as the string multiplets of the type IIA 
theory in eight dimensions.

Now, to obtain the classical solutions including their
energies in terms of the 
relevant quantum numbers, we also write down the 
Hamiltonian corresponding to the  classical action 
(\ref{classical-action}) and observe
an $SL(3)$ symmetry in this case as well.
Starting from the action in eqn.(\ref{classical-action}), 
the canonical momenta for the gauge field components $A_i$ are: 
\begin{equation}
\Pi_i \equiv {\partial {\cal L}\over \partial{(\partial_t A_i)}}
= {1\over g}F_{t i} - 2  \epsilon^{m n}\epsilon^{i j}
        C_m \partial_j \phi_n,
\end{equation}
and $\Pi_0 = 0$. The canonical momenta for the scalar 
fields $\phi_m$ in three dimensions are:
\begin{equation}
P_m \equiv {\partial {\cal L}\over \partial{(\partial_t \phi_m)}}
= {1\over g} g^{m n} \partial_t \phi_n -  \epsilon^{m n}
        \epsilon^{i j}C_n F_{i j}.  \label{canoP}
\end{equation}
We will consider the classical solutions in three dimensions without
magnetic charges, as a result the last term in eqn.(\ref{canoP}) is absent.
Since $\Pi_0 = 0$ , the conservation equation for 
momenta $\Pi_{\mu}$ gives,
\begin{equation}
\partial_i \Pi_i = 0,           \label{gauss}
\end{equation}
also known as the Gauss-law constraint. 
This equation can be solved by defining $\Pi_i = \partial_i \Lambda$,
where $\Lambda$ satisfies  
$\partial^i \partial_i \Lambda = 0$. It can then also 
be shown that, for static configurations,
the $\phi_m$ equation of motion simplifies to 
$\tilde{g}^{m n}\partial^i\partial_i \phi_n = 0$,
with $\tilde{g}^{m n} = g^{m n} +  
2 \epsilon^{m p}C_p \epsilon^{n q} C_q g^2$. 
The modified metric $\tilde{g}^{m n}$ coincides with the one
appearing as components of ${\cal M}$  
in eqn.(\ref{sl3-matrix}) and corresponds to the fact that the scalars
$\phi_m$ couple to an appropriate form of metric 
required by an $SL(3)$
invariance. The form of the Hamiltonian is given by an expression:
${\cal H} = H + H_S$, where 
\begin{equation}
H = {g \over 2} (\Pi_i + 2 \epsilon^{m n}\epsilon^{i j}
C_m \partial_j \phi_n)^2 + {1\over 2 g } g^{m n} 
\partial_i \phi_m \partial_i \phi_n,   \label{hamiltonian}
\end{equation}
and $H_S = {1\over 2 g}(\partial_i X^I)^2$ is the Hamiltonian for the scalars
$X^I$.

We will now obtain classical configurations which are the 
BPS solutions for the Hamiltonian (\ref{hamiltonian}) and identify 
their energies with
the tensions of the $(p, q, r)$-strings in eight dimensions. 
The string tensions for the eight-dimensional strings 
are given in terms of quantum numbers $ p_a (a=1,..,3) = (p, q, r)$,
as well as the $SL(3)/SO(3)$ moduli and is expressed by the 
length of a vector $X_i$ defined in terms of the moduli and
the quantum numbers $(p, q, r)$ as:  
\begin{equation} 
        X_i = ({\lambda_0}^{-1})_{i a} p_a
                                \label{charge}
\end{equation}
where $\lambda$ are the ``vielbein's'' satisfying 
$\lambda \lambda^T = {\cal M}$. Explicitly, we have:
\begin{eqnarray}
        \lambda^{-1} = \pmatrix{ e^{-{(\phi + \alpha)/2}}
        & - \chi e^{- {(\phi + \alpha)/2}} &
        - e^{- {(\phi + \alpha)/2}} a_1 + 
        \chi e^{- {(\phi + \alpha)/2}} a_2      \cr
        0 & e^{\alpha/2} & - e^{\alpha/2}a_2 \cr
        0 & 0 & e^{\phi/2} }.
                \label{lambda}
\end{eqnarray}
For a special case when the
moduli $\chi = a_1 = a_2 = 0$, 
string tension for the $(p, q, r)$-string is given by\cite{bhatt,luroy} 
\begin{equation}
T(p, q, r)  = \left[e^{-(\phi + \alpha)} p + e^{\alpha} q
                + e^{\phi} r \right]^{1\over 2}
            = \left[ p + e^{\phi + 
                2\alpha} q + e^{2\phi + \alpha} r \right]^{1\over 2}
                T_{(1, 0, 0)}.          \label{tension}
\end{equation}

Now, the solution for a $(1, 0, 0)$ string is simply obtained from 
the equation of motion for $\Lambda$ in the presence of a source
term. The addition of the source term is a modification of the 
(Gauss-law) constraint equation (\ref{gauss}) in presence of
charges deposited by the strings attached
to the branes. In the worldvolume theory we are studying, such
source terms come from the coupling terms involving various 
worldvolume and space-time fields and is in fact finally
responsible for the appropriate $SL(3)$ multiplet structure of the
classical solutions we are discussing. 
In our case, the presence of a source is also seen by the fact that the 
BPS condition requires the turning on of one of the scalar fields,
say $X^9$, at the same time $\Lambda$ is turned on. 
Similarly, the solutions of the Green function equation for 
fields $\phi_1$ and $\phi_2$ will be interpreted as $(0, 1, 0)$ and
$(0, 0, 1)$ strings. Once again appropriate scalars are turned on to 
maintain the BPS nature.

To discuss the supersymmetry of the classical solution, we 
now write down the world-brane action in the presence of 
fermions. It is once again possible to consider
the full $SL(3)$ invariant action in presence of
a general background. However, to match with the special set 
of string tensions given in (\ref{tension}),  
it is enough to consider the action with
zero background and trivial off-diagonal moduli. 
In this case, our action (before dualization)
corresponds to 
a straightforward compactification and reduction
of ten dimensional super Yang-Mills
giving rise to the following action:
\begin{equation}
S = {1\over g} \int d^3x \left[-\frac{1}{4} (F_{\mu\nu})^2 
-\frac{1}{2} (\partial_\mu\phi_m)^2 
-\frac{1}{2} (\partial_\mu X_I)^2 
+ \frac{i}{2} \bar\psi^{(a)} \gamma^\mu\partial_\mu\psi^{(a)}
\right],
\label{action} \end{equation}
where ten-dimensional
spinors were splitted into 8 three-dimensional
spinors $\psi^{(a)}$,
 $a = 1...8$ and each of the $\psi^{(a)}$ is
the 2-spinor in three dimension.

The action has an $N=8$ supersymmetry in three dimension.
The supersymmetry transformations follow directly
from the ten dimensional theory:
\begin{equation}
\delta \psi^{(a)}_\alpha = F_{\mu\nu}
\gamma^{\mu\nu}\epsilon^{(a)} +
2i (\partial_\mu\phi_m) \gamma^\mu(\Sigma_m)^{aa'}
\epsilon^{a'}  +
2i (\partial_\mu X_I)\gamma^\mu(\Sigma_I)^{aa'}
\epsilon^{a'}a, \label{susy}
\end{equation}
where $(\Sigma_m, \Sigma_I)$ are the
$SO(7)$ gamma matrices,
and the 8 spinors $\epsilon^{(a)}$'s correspond
to 8 supersymmetry transformations.

In order to construct the one-half BPS configurations
in this world volume theory which correspond
to the end of a $(p, q, r)$-string in eight dimensions
we will start with the $(1,0,0)$ case. The fact
that the 2-form field corresponding to this 
charge couple to $F_{\mu\nu}$ in the Chern-Simmons
term, implies that the end will deposit an electric charge.
Since the two compact scalars $(\phi_m)$ are equivalent 
to the electromagnetic field, one can also consider
the configurations with ``electric charges''
associated with them. As we will see 
these configurations are characterized
by non-trivial windings of the compact
scalars. From the coupling of these scalars
with the other two 2-form fields one can identify them
with the end of $(0,1,0)$ and $(0,0,1)$
strings respectively.

Let us now start with the configuration with
an electric charge. The field strength
becomes $F_{0r} = \frac{c}{r}$ where 
we are using polar coordinates for space $(r,\theta)$, and
other components of the field strength
vanish. This configuration does not preserve
any supersymmetry. However one can excite 
a higgs field $X^9$ simultaneously \cite{calmal}
such that $F_{0r} = \partial_r X_9$. Setting
all other fields $\phi_i$ and $X_I$ to zero
and substituting in (\ref{susy}) one can show 
explicitly that it projects out one half 
supersymmetry given by
\begin{equation}
\delta\psi = 0 \Rightarrow
[ 1 + i\gamma^0\Sigma^9 ]\epsilon = 0.  
\label{susy1} \end{equation}
A similar condition is also satisfied when 
an F-string in ten dimensions ends on the D3-brane:
$(\Gamma^{0r} + \Gamma^{r9})\epsilon =0$,
which was seen in the case of $N=4$ gauge theory  
in four dimensions. Clearly the above condition will project out
one half supersymmetry.


The excitation of the Higgs field 
implies that the configuration of
the brane in the transverse space 
gets modified due to the string.
The $\log{r}$ dependence of $X^9$
near $r=0$ can be interpreted as
near the position of the electric 
charge, due to the pulling of the
string the brane gets projected
in a direction transverse to its
world volume forming a spike like
configuration\cite{hashi1,hashi2}. 
That, this is really the U2-brane 
gets projected instead of the string 
itself, can be argued on the ground 
that locally the spike does not have
any B-field charge\cite{hashi1}.
Unlike the 3-brane case, the behaviour 
of the higgs far from the electric 
charge also has a $\log{r}$ divergence.
This is associated with the 
characteristic infra-red divergence 
of gauge theories in 3 dimensional 
space time and has already been 
discussed. 

Now consider the $(0,1,0)$ string which
can be thought of as the D-string in the eight 
dimensional compactified theory. In the 
case of D3-brane we know that the end of
a D-string deposits a magnetic charge
\cite{calmal}.
However, the non-zero magnetic charge
configuration in four dimension 
$F_{\theta\phi} \ne 0$
correspond to a non-zero winding of one 
compact scalar on the spatial boundary. As an example we 
associate it with the generator 
$\Sigma_3$ in eqn.(\ref{susy}).
Just like the electric
charge case, this will break all the supersymmetries
and one has to excite a higgs field $X_8$
satisfying $\partial_r X^8 = \frac{1}{r}
\partial_\theta\phi$ with 
$\partial_\theta X_8 = 0$. 
Substituting this configuration in 
eqn. (\ref{susy}) one gets
\begin{equation}
\delta\psi = 0 \Rightarrow
[1+\gamma^0\Sigma^3\Sigma^8]\epsilon = 0 , 
\label{susy2} \end{equation}
which once again projects out one-half supersymmetry.
A similar configuration will provide
the $(0,0,1)$ case as well. Note that there is
no direct analogue of the $(0, 0, 1)$ case in four dimensions.
In fact the associated string correspond 
to wrapped D3 brane in ten dimension.

To compare the energy of the $1/2$ BPS states
just discussed with the string-tension given in 
eqn.(\ref{tension}), we evaluate the Hamiltonian derived in 
eqn.(\ref{hamiltonian}). Once again, restricting oneself to the 
static configurations and setting the off-diagonal couplings in 
(\ref{sl3-matrix}) to zero, for simplicity, we have:
\begin{equation}
H = {1\over g} \int d^2r \left[\frac{1}{2}E^2 
+ \frac{1}{2}(\nabla\phi_m )^2 
+ \frac{1}{2}(\nabla X_I)^2 \right],
\label{energy}\end{equation}
where in addition we have assumed that the three-dimensional magnetic 
charges are absent for our solutions. 
Substituting the configuration 
$F_{0 r} = \partial_r X_9$ we now get the energy-expression 
for $(1,0,0)$ charge:
$E = {2\pi c^2\over g} log(\frac{R}{\delta})$.
After a normalization of $c$ using flux quantization,
one gets\cite{calmal} $E = T_f X^9$, and 
$T_f = {1\over 2\pi}$ is the string tension of a $(1, 0, 0)$ string. 
The integral (\ref{energy}) diverges at both the limits
and so we use $\delta$ and $R$ as the UV
and IR cut-off. The divergence as one approaches
near the electric charge implies
that the spike is infinitely long. The other
cut-off is a charachteristic of 3-dimensions
and is originating from the infinite volume
of the U2-brane. We have already discussed this issue and its
possible remedies earlier. However, it is interesting to observe that
$X^9$ interpreted as a string coodinate also 
has similar divergence. As a result one still has the energy 
proportional to the length of the string coordinate. Similarly the 1/2 BPS 
solution given by the conditions: $\partial_r X^{8, 7}$ = 
${1\over r}\partial_{\theta}\phi_{1, 2} = {c\over r}$ match with the 
$(0, 1, 0)$ and $(0, 0, 1)$ solutions respectively,
provided the constant $c$ in these cases are again appropriately 
normalized\cite{calmal}. As stated earlier, the relative 
normalizations of the constants appearing 
in the three cases follow from the
$SL(3)$ invariance of the coupling terms involving the worldvolume 
coordinates and the background space-time fields. Alternatively
these normalizations also follow from the charge
quantizations in the world-volume theory. In other words the 
$SL(3)$ symmetry in the worlvoume theory gives appropriate 
duality multiplets in the space-time theory.  
The stucture of the Hamiltonian in eqns.(\ref{hamiltonian})
and (\ref{energy}) implies the
matching of the tension with the classical energy of the solutions
hold in the presence of other moduli as well. 


We have therefore shown the presence of string 
multiplets in a world-volume theory obtained from the 
compactification of a D4-brane theory. The key factor in this study 
has of course been the $SL(3)$ invariance of the compactified 
D4-brane action which in turn follows from its M5-brane interpretation. 
We have studied this action in the linear limit, 
${\alpha}^{\prime} \rightarrow 0$, in which the higher derivative terms coming 
from the Born-Infeld action drop out.

\def\E{\vec E}
\def\N{\vec\nabla}

However, since the dynamics of the D-brane action is described
by the Born-Infeld (B-I) action \cite{bachas} it is necessary to 
extend the previous analysis of linear action to the 
B-I case also. Our main goal is to check whether the
solutions that we obtained are valid in the non-linear 
case also. That they are BPS can be checked by taking
the supersymmetry into account though we will confine
to the bosonic part of the action. We will also  observe that the 
relaxation of  the BPS condition gives a large number of 
other solutions even in the static case. However we retsrict 
ourselves to the case of spike solutions only.

The dynamics of a pure D-brane of dimension p is described
by the action which can be obtained from the
$U(1)$ B-I action in $(9+1)$-dimensions reduced
to $(p+1)$ dimensions. Since the branes we are considering are
obtained by compactifying the D-branes on a
torus (in our case) the corresponding action
can be obtained by compactifying the B-I action
for D-brane on a torus as in the linear case.
We will use the Hamiltonian
formulation as that is more suitable and
also, for the sake of simplicity, we will
consider the reduced action obtained by exciting
only those fields which are relevant for the spike 
configuration and setting other fields to zero.
That this is a valid configuration can be checked
from the equations of motion.

Born-Infeld action in ten dimension is given by
\begin{equation}
L = -\frac{1}{g} \int d^3x \sqrt{-\det(g + F)},
\end{equation}
where $g_{MN}$ is the metric tensor and $F_{MN}$ is 
the $U(1)$ electromagnetic field. 
The B-I action for a p-brane can be obtaines by identifying
$A_M = A_{\bar{\mu}}$ for $M = 0,1,..,p$ and
$A_M = X_I$ for $I = p+1,...,9$.
For obtaining the U2-brane in 8-diemnsions
one should also identify fields as before:
$A_M = A_\mu$ for $M = 0,1,2$ and
$A_M = \phi_m$ for $M = 3,4;$ and 
$A_M = X_I$ for $M = 5,...,9$. 
We will moreover restrict 
ourselves to the flat Euclidean space-time.

To start with we consider the configuration containing
electric charge deposited on the brane worldvolume.
That will give a radial electric field. As we saw 
earlier we have to excite a non-compact higgs field
also to make it BPS. So we consider the reduced action 
with $X\ne 0$,
$\E \ne 0$, where $E_i = F_{0i}$ is given
by
\begin{eqnarray}
L &=& - \frac{1}{g} \int d^3x v , 
\quad\quad\quad {\mathrm where} 
\nonumber  \\
v &=& \sqrt{(1-\E^2)(1+(\N X)^2) + (\E.\N X)^2 - {\dot X}^2} 
\end{eqnarray}
This case was studied in \cite{calmal} for D-3 brane
worldvolume and  the analysis is similar.
The momenta are given by
\begin{equation}
g \vec\Pi = \frac{\vec E(1 + (\N X)^2) - \N X(\E.\N X)}{v}, 
\quad
g P = \frac{\dot X}{v}. 
\end{equation}
The Hamiltonian can be written as 
\begin{equation}
H = \frac{1}{g} \int d^2x 
\sqrt{(1 + (\N X)^2)(1+g^2P^2) + g^2(\Pi^2 
+ (\vec \Pi.\N X)^2 )}.
\end{equation}
We also have to take care of the Gauss's law
constraint given by $\nabla . \Pi = 0$.

We can solve the Gauss's law by setting
$\vec\Pi = \N\Lambda$ with ${\nabla}^2\Lambda 
= 0$. To get the higgs field configuration
we consider the static solution obtained by
setting $P = 0$ and $\dot P = 0$. That will lead to
the equations
\begin{equation}
\nabla . [\frac{\N X + g^2\vec\Pi(\vec\Pi.\N X))}
{\sqrt{1 + (\N X)^2 + g^2(\Pi^2 
+ (\vec \Pi.\N X)^2 )} }]  = 0.
\end{equation}
If one imposes the condition, $g\vec\Pi = \N X$ 
the above condition gets reduced to the Gauss's
law $\N . \vec\Pi = 0$. 
The energy can be obtained by substituting
the solution in the Hamiltonian and leads to
\begin{equation}
H = \int d^2x [ 1 + |\N\Lambda |^2 ]
\end{equation}
The argument in the square root comes out
to be perfect square to make the energy
expression linear, as is expected for a
BPS solution. Moreover the energy of the 
solution matches exactly with that of the
linearized theory.
This equation also has other solutions 
as mentioned in \cite{calmal}. However 
we do not discuss them here.

Now we look for the BPS configuration using the 
compact scalars $\phi_m$. For that purpose
we write down the reduced action containing
a compact scalar $\phi$ and a non-compact higgs
$X$ setting all other fields to zero.
Actually this is equivalent to the above
case due to the fact that the electromagnetic 
field is equivalent to a compact scalar in (2+1)
dimension.

For this configuration, the action
can be written as
\begin{eqnarray}
L &=& - \frac{1}{g} \int d^3x u
\quad\quad\quad {\mathrm where}
\nonumber  \\
u &=&
\sqrt{[1-(\nabla_0\phi)^2 +(\N\phi)^2]
[1-(\nabla_0 X)^2 +(\N X)^2] 
- [\nabla_0\phi.\nabla_0 X - (\N X.\N\phi)]^2} \cr
 & &
\end{eqnarray}

For the convenience of calculation, we
introduce a $2\times2$ matrix given by
\begin{equation}
 M = \left(\begin{array}{cc}
1+|\N X|^2 & -(\N X.\N\phi) \\
-(\N X.\N\phi) & 1+|\N\phi|^2
\end{array}\right).
\end{equation}
In terms of this matrix $u$ can
be written as,
\begin{equation}
u = \sqrt{ \det{M} - 
\left(\begin{array}{cc}
(\nabla_0\phi & \nabla_0 X) 
\end{array}\right) 
 M 
\left(\begin{array}{c}
\nabla_0\phi \\ \nabla_0 X \end{array}
\right)}, \end{equation}
and the momenta can be obtained from the
above action as:
\begin{equation}
\left(\begin{array}{c}P_\phi \\ 
P_x \end{array}\right)
 = \frac{1}{gu} ( M )
\left(\begin{array}{c}\nabla_0\phi \\ 
\nabla_0 X \end{array}\right).
\end{equation}

The Hamiltonian is given by
\begin{equation}
H = \frac{1}{g} \int d^2x \sqrt{\det{M}
(1 + g^2 \tilde P M^{-1} P )},
\end{equation}
where $P$ is the 2 component column
vector $\left(\begin{array}{c}P_\phi 
\\ P_x \end{array}\right)$.

We are interested in the static, zero momentum
solution. 
Setting the time derivatives of the momenta to
zero, with $P=0$ we get the equations of motion
\begin{eqnarray}
\dot P_\phi &=& \partial_m \left[{{\partial_m \phi + 
(\N\phi\times\N X)\epsilon_{mn}\partial_n X }
\over {\sqrt{det M}}}\right] = 0, \\
\dot P_x &=& \partial_m \left[{{ \partial_m X - 
(\N\phi\times\N X)\epsilon_{mn}\partial_n \phi]}
\over{\sqrt{det M}}}\right] = 0,
\end{eqnarray}
where $m, n = 1,2$ are the space indices, and
the cross product of two spatial vectors is given by
$\vec A\times\vec B = \epsilon_{mn}A_mB_n$.

These equations have many solutions
but we are interested in the vortex-type
solution only which is given by.
\begin{equation}
\partial_\theta\phi = r\partial_r X = n ,\quad 
\partial_\theta X = \partial_r\phi = 0.  
\end{equation}
Clearly this configuration satisfies the equations
for constant value of $n$. This is the number
representing the winding of the compact scalar
about the origin of two dimensional space. Substituting 
this solution in the expression of the Hamltonian
one again gets the energy 
\begin{equation}
E = \int d^2x [1 + |\N X|^2].
\end{equation}
Interestingly, like the electric charge 
case, the nonlinear form of the B-I energy
gets linearized for this configuration also.
Similar kind of linearizations of BI energy
have been discussed in\cite{brech}.

We have therefore shown the existence of the 
``spike solutuions'' in the case of B-I action also. 
Each of them correspond to
a singularity on the 2-space of worldvolume and represents
the end of the string. 
Finally, there are other 
interesting soluitions also like brane-antibrane which can be
be studied in this formulation.

Our investigation so far has been 
restricted to the 1/2 BPS 
solutions of the three dimensional 
theory. We now discuss the 
possibility of other BPS states. 
For this purpose we consider the
non-abelian gauge theory. Unlike the 1/2
BPS states, the BPS states with less
supersymmetry requires more than one
"electric fields" and the 1/4 and 1/8
BPS configurations are associated with
two or three non-parallel charges respectively
which are similar to and a generalization
of the dyonic configurations in the four
dimensions. In this discussion we will
restrict ourselves to the static configurations
with zero magnetic field. 

The non-abelian generalization 
of the worldvolume theory
arises from the presence of multi D4-branes 
in their compactified
form. The associated gauge theory Hamiltonian 
in the linearized version is now written as:
\begin{eqnarray}
{\cal H} &=& \frac{1}{2 g} \int d^2x Tr[
E^2 + B^2 + (D_0\phi_m)^2 + (D\phi_m)^2
\nonumber  \\
&+& (D_0 X^I)^2 + (D X^I)^2 + {\mathrm commutators} ]
\end{eqnarray}
where electric field $E$ is a 2-vector  
and $B$ is a scalar
with respect to the spatial rotation group.
The covariant derivative is
written as $D_\mu Y = \partial_\mu Y -i [ A_\mu, Y]$
for any field $Y$ in the adjoint of $SU(N)$. 
$D\phi_m$ and $D X^I$ are the 
covariant derivatives with respect to the 
two spatial directions. 

To discuss the BPS bounds on solutions 
one introduces certain parameters, 
which are angles among the charges, 
into the Hamiltonian  and 
extremization of the action with 
respect to these paramaters
then gives the expression for the 
BPS saturated energy. 
Situation with $1/4$ BPS states can 
be discussed in the three
dimensional case following the results in four 
dimensions\cite{hashi2,oku,lee}. 
We now examine the possibility 
of $1/8$ BPS states in the three
dimensional theory. In our
case, since we need configurations
with three non-parallel charges, 
we introduce a 3-dimensional 
rotation among the 
dynamical variables.  
Also following \cite{hashi2}, we note
that certain  asymptotic solutions 
having the prong-structure
in the three dimensional gauge 
theory do satisfy the 
conditions we obtain below. 
This suggests the identification
of the non-planar networks with these
configurations. As in \cite{hashi2}
we expect such solutions to be 
modified at short distances. 
These asymptotic solutions are 
essentially the superposition of 
the ones for the individual branes.

Introducing new angles, we rewrite 
the Higgses as 
$\tilde X_I = R_{IJ} 
X_J$ for $I, J = 7,8,9$ and $R_{IJ}$ is 
the $SO(3)$ rotation matrix. Now with a little
algebra the Hamiltonian can be written
as 

\begin{eqnarray}
H &=& {1\over g} \int d^2x \frac{1}{2} Tr \{
[ (E_i-D_i\tilde X_7)^2 + 
(\epsilon_{i j}D_j \phi_1 - D_i\tilde X_8)^2 +
(\epsilon_{i j} D_j \phi_2 - D_i\tilde X_9)^2]
\nonumber \\ &+&
2 [E_iD_i\tilde X_7 + 
\epsilon_{ij}D_i\phi_1 D_j \tilde X_8 +
\epsilon_{ij}D_i\phi_2 D_j \tilde X_9 ] \},
\end{eqnarray}
where we have suppressed the time derivative
and the commutator terms.


Let us define the component of the charges: 
\begin{equation}
Q_1^1 = \int tr(E_i D_i X_7), \quad
Q_2^2 = \int tr(\epsilon_{ij}D_i\phi_1 D_j X_8), \quad
Q_3^3 = \int tr(\epsilon_{ij}D_i\phi_2 D_j X_9), 
\end{equation}
and six others obtained by permuting the Higgs fields
in the above equation. So each of the charge
is a vector in the three diemnsional space. These
charges are essentially generalization
of the electric and magnetic charges
in four diemnsions. However, it remains 
to see, whether these charges appear as
the central charges of the $N=8$ supersymmetry
algebra.

Following \cite{fraser,hashi2}, the BPS bound can be
written from the Hamiltonian as
\begin{equation}
E_0 = \sum_{I, J} R_{I J}(\theta, \phi, \psi) Q_I^J,
\label{min}\end{equation}
where $\theta$, $\phi$ and $\psi$ are the
Eulerian angles and in order to get the
bound one has to maximize (\ref{min}) 
with respect to the angles. 

The equations for which the bound is satisfied are given
by the vanishing of the squares and commutators. In addition, there
are two other constraints, the Gauss's law and the fact that
$\phi$'s take values in the compact space. So we 
get equations:
\begin{eqnarray}
\quad E - D\tilde X_7 \quad &=& 0 \\ 
\epsilon_{ij}D_j \phi_1 - D_i \tilde X_8 &=& 0 \\
\epsilon_{ij}D_j \phi_2 - D_i \tilde X_9 &=& 0
\end{eqnarray} 
with $F_{ij} = 0$, $D_0 X_I = 0$, $D_0\phi_a = 0$
and the commutators are vanishing.

Clearly, one can see that the one-half
BPS configurations are solutions of
these equations in the assymptotic
region. Near origin the solution
should be regular. 
However it remains to be seen whether, like the 
$1/4$ supersymmetric case in four dimensions, these 
conditions are also 
sufficient to determine the angles 
appearing in eqn. (\ref{min}) 
uniquely and in turn determine the BPS bound. 
As mentioned earlier, like the four dimensional 
case\cite{hashi2},
one can show that a multi-prong planar structure 
in three dimensions 
provides an asymptotic form for solving these equations, 
the simplest possibility now being a 4-prong structure 
meeting at a point in a three dimensional space.  
Moreover the orientation of the legs in the prong also 
coincides with the ones expected from the U-duality and 
supersymmetry conditions. This construction gives further 
evidence that string multiplets of the eight-dimensional 
type II theory and their networks can have interesting role to 
play in certain three dimensional gauge theories.
It is however left as a future exercise to 
solve the BPS conditions completely.

\section{Acknowledgements}
We like to thank Carlo Angelantonj, Ignatios Antoniadis, 
Costas Bachas, Massimo Bianchi, Keshav Dasgupta, 
Elias Kiritsis, Sunil Mukhi 
and Koushik Ray for invaluable discussions at various stages of 
the work. A.K. also thanks Centre de Physique Theorique, Ecole
Polytechnique and all its members 
for the warm hospitality during the course of this work. 

\vfil
\eject


\begin{thebibliography}{99}
\newcommand{\np}{{\it Nucl. Phys.} {\bf B}}
\newcommand{\pl}{{\it Phys. Lett.} {\bf B}}
\newcommand{\prd}{{\it Phys. Rev. }{\bf D}}
\newcommand{\prl}{{\it Phys. Rev. Lett.}}
\newcommand{\mpl}{{\it Mod. Phys. Lett. }{\bf A}}
\newcommand{\ijmp}{{\it Int. J. Mod. Phys.}{\bf A}}
\newcommand{\cqg}{\it Class. Quant. Grav.}
\newcommand{\jmp}{\it J. Math Phys.}
\newcommand{\cmp}{\it Comm. Math. Phys.}

\bibitem{calmal} C. Callan and J. Maldacena, ``Brane Dynamics from the
Born-Infeld Action'', {\np}{\bf 513} (1998)
198; hep-th/9708147.

\bibitem{gibbons} G. Gibbons, ``Born-Infeld Particles and Dirichlet 
p-branes'', {\np}{\bf 514} (1998) 603; 
hep-th/9709027. 
 
\bibitem{hashi1} A. Hashimoto, ``The Shape of Branes Pulled by
Strings'', {\prd}{\bf 57} (1998) 6441; hep-th/9711097. 



\bibitem{gauntlett} J. Gauntlett, J. Gomis and P. Townsend, 
JHEP {\bf 01} (1998) 003; ``BPS Bounds for Worldvolume Branes'',
hep-th/9711205.

\bibitem{prasanta} P. Tripathy, ``Self-Dual Maxwell Chern-Simons 
Solitons in 1+1 Dimensions'', hep-th/9811186. 



\bibitem{bachas} J. Polchinksi, ``TASI lectures on D-branes'',
hep-th/9611050; C. Bachas, ``Lectures on D-Branes'', hep-th/9806199.

\bibitem{cal2} C. Callan, A. Guijosa and G. Savvidy, ``Baryons and
String Creation from the Fivebrane Worldvolume Action'', 
hep-th/9810092.


\bibitem{bergmann} O. Bergmann, ``Three-Pronged Strings and 
1/4 BPS States in N=4 Super Yang-Mills Theory'', hep-th/9712211;
O. Bergmann and A. Fayyazuddin, hep-th/9802033. 

\bibitem{net} A. Sen, ``String Network'', hep-th/9711130; 

\bibitem{net2} J. Schwarz, hep-th/9607201; 
M. Gabardiel and B. Zwiebach, ``Exceptional 
Groups from Open Strings'', hep-th/9709013;
K. Dasgupta and S. Mukhi, ``BPS Nature of 
3-String Junctions'', hep-th/9711094;
Y. Matsuo and K. Okuyama, hep-th/9712070;
C. Callan and L. Thorlacius, {\np}{\bf 534} (1998) 121. 

\bibitem{sl2} J. Schwarz, {\pl} {\bf 360} (1995) 13, [hep-th/9508143].


\bibitem{hashi2} K. Hashimoto, H. Hata and N. Sasakura, 
``3-String Junction and BPS Saturated Solutions in SU(3)
Supersymmetric Yang-Mills Theory'', {\pl}{\bf 431} (1998) 303;
hep-th/9803127; ``Multi-Pronged Strings and BPS Saturated Solutions in 
SU(N) Supersymmetric Yang-Mills Theory'', {\np}{\bf 535} (1998) 83;
hep-th/9804164. 

\bibitem{oku} T. Kawano and K. Okuyama, ``String Network and
1/4 BPS States in N=4 SU(n) Supersymmetric Yang-Mills Theory'',
{\pl}{\bf 432} (1998) 338; hep-th/9804139. 

\bibitem{lee} K. Lee and P. Yi, ``Dyons in N=4 Supersymmetric Theories
and Three-Pronged Strings'', {\prd}{\bf 58} (1998) 066005;
hep-th/9804174.


\bibitem{bhatt} S. Bhattacharyya, A. Kumar and S. Mukhopadhyay,
``String Network and U-Duality'', {\prl} {\bf 82} (1998) 754;
hep-th/9801141; A. Kumar and S. Mukhopadhyay, " Supersymmetry and U-brane Networks", hep-th/9806126 ( to appear in Int. Jour. Mod. Phys. A )

\bibitem{leung} N. Leung and C. Vafa, {\it Adv. Theor. Math. Phy.}
{\bf 2} (1998) 91; hep-th/9711013. 

\bibitem{luroy} J. Lu and S. Roy, ``U-Duality p-Branes in Diverse 
Dimensions'', hep-th/9805180; ``4-String Junction and its 
Network'', hep-th/9807139. 

\bibitem{christof} C. Schmidhuber, ``D-Brane Actions'',
hep-th/9601003; A. Tseytlin, ``Self-Duality of Born-Infeld
Action and Dirichlet 3-brane of type IIB Superstring Theory'',
{\np} {\bf 469} (1996) 51;M. Abou Zeid and C. Hull, ``Intrinsic Geometry of 
D-Branes'', hep-th/9704021; 
P. Townsend, ``Membrane tension and manifest IIB
S-Duality'', hep-th/9705160;
M. Caderwall and P. Townsend, ``The Manifestly
$SL(2, Z)$-Covariant Superstring'', JHEP {\bf 09} (1997)
003; hep-th/9709002; S. Mukherji, ``On the SL(2, Z)-Covariant Worldsheet
Action with Sources'', hep-th/9805031.





\bibitem{taylor} W. Taylor IV, ``D-Brane Field Theory on 
Compact Spaces'', {\pl}{\bf 394} (1997) 283. 


\bibitem{verl2} F. Hacquebord and H. Verlinde, {\np}{\bf 508}
(1997) 609; C. Hofman, E. Verlinde and G. Zwart, ``U-Duality 
Invariance of Four-Dimensional Born-Infeld Theory'', hep-th/9808128; 
C. Hofman and E. Verlinde, ``U-Duality of Born-Infeld
on the Non-Commutative Two-Torus'', hep-th/9810116. 

\bibitem{sen3d} A. Sen, ``Strong-Weak coupling Duality in three
dimensional string theory'', {\np}{\bf 434} (1995) 179. 

\bibitem{seiberg} N. Seiberg and E. Witten, ``Gauge Dynamics and 
Compactification to Three Dimensions'', hep-th/9607163; 
N. Seiberg, ``Notes on Theories with 16 Supercharges'', hep-th/9705117. 


\bibitem{vafa} C. Vafa, {\np}{\bf 469} (1996) 403 [hep-th/9602022].


\bibitem{fraser} C. Fraser and T. Hollowood, ``Semi-classical 
Quantization of N=4 supersymmetric Yang-Mills Theory and 
Duality'', {\pl}{\bf 402} (1997) 106.  

\bibitem{brech} D. Brecher, ``BPS states of the 
nonabelian Born-Infeld action '', hep-th/9804180;  
D. Brecher and M. J. Perry,
"Bound states of D-branes and the
non-abelian Born-Infeld action"
{\np}{\bf 527} (1998) 121 [hep-th/9801127].

\end{thebibliography}
\end{document}